\documentstyle[12pt,blois]{article}

\begin{document}
\heading{PHOTOMETRIC REDSHIFT TECHNIQUES: \footnote
{To appear in the proceedings of the Xth Rencontres de Blois:
{\it The Birth of Galaxies}} \\
      RELIABILITY AND APPLICATIONS}

\author{H.~K.~C.~Yee } 
{ Department of Astronomy, University of Toronto, Toronto, M5S 3H8, Canada.}  
{~}

\begin{bloisabstract}
The idea of using multi-band photometry to estimate the redshifts of galaxies
has a long history, but it is only recently that it has come
to widespread use and acceptance.
This paper describes the recent developments and applications
of photometric redshift techniques.
There are two major approaches to photometric redshift techniques:
spectral energy distribution (SED) fitting, using either
observed  SEDs or population synthesis model SEDs; and
empirical methods using spectroscopic samples as training sets.
The availability of Hubble Deep Field data in four bands
has provided a great impetus in using photometric redshifts.
A detailed discussion of the robustness of the results is 
presented.
The usefulness of redshift information derived from 
photometric data with limited number of filters (as few as two)
is also discussed.
Taking all current evidence together, it appears that photometric
redshift can be a reliable and extremely useful technique with many
applications, if care is taken to ensure that systematic
effects and catastrophic redshift errors are minimized.

\end{bloisabstract}

\section{Introduction}

The idea of using broad-band photometry to estimate the redshifts of
galaxies was suggested as early as 35 years ago by Baum (1963).
The basic concept is simple: multi-band photometry can be regarded
as a very low resolution spectrum, and hence can be used
to estimate the redshifts of galaxies and other extragalactic objects.
Although this idea has been applied by a number of investigators
in the ensuing years, it was only in the past few years that photometric
redshifts have found widespread application and acceptance.

With the availability of deep, wide-field CCD photometry today,
the potential scientific return of photometric redshifts is great.
There are essentially two major applications of the method.
The first is to provide redshift samples in a quick and inexpensive way.
Although the typical redshift uncertainty is expected to be in
the range of 0.05 to 0.10, far less accurate than that expected
from a spectroscopic survey, many scientific goals can still
be attained with a photometric redshift sample; for examples:
the determination of luminosity functions (e.g., SubbaRao et al. 1996;
Mobasher et al. 1996, Gwyn \& Hartwick 1996; 
Sawicki, Lin, \& Yee 1997, hereafter SLY97), 
luminosity densities (e.g., SLY97; Connolly et al. 1997; Madau et al. 1998);
and the projected spatial correlation function  (e.g., Connolly, Szalay, \&
Brunner 1998; Miralles \& Pell\'o 1998). 
The second major application is to provide estimates  of 
the redshifts of objects which
are too faint for conventional spectroscopy.

In this review, I will begin with some historical notes on the method
of photometric redshifts, describe the recent developments, and comment
on the reliability and recent applications of the method.
The technique of using simply  two or three color bands for 
selecting a sample of objects at a certain redshift, which
is the photometric redshift method in its crudest form,
 will also be discussed.

\section{Photometric Redshift Techniques}

\subsection{Some Historical Notes}

The first systematic application of photometric redshift was carried
out by Baum (1963) who obtained photoelectric photometry of early-type
 galaxies in distant galaxy clusters  using 9 bands.
He estimated the redshifts of the clusters by comparing the spectral
energy distributions (SEDs) with those from galaxies in the Virgo Cluster,
using primarily the position of the 4000\AA~break in early-type galaxies.

Koo (1985) analysed the dependence of 4-band data
 from photographic photometry on redshift.
By making linear combinations of the 4 filters, he was able to show
that one can form ``iso-$z$'' loci on color-color plots, and hence
provide a ``poor-person's redshift machine''.

A strong effort in applying photometric redshifts to a specific astronomical
problem was first made by Loh \& Spillar (1986).
They used CCD photometry in 6 filters and obtained redshifts
by fitting galaxy SEDs of various morphological types, 
and attempted to determine
$\Omega$ via galaxy number density as a function of redshift.

Many of the early efforts, however, were hampered by the lack of
high quality, wide-field digital data, and the difficulties of 
obtaining a significant and convincing set of verification redshifts.

\subsection{Modern Developments}

After these early efforts,
there was a hiatus of almost 10 years
before photometric redshifts came into vogue again.
There are several reasons why photometric redshifts have enjoyed such a
rigorous revival.
One is the advent of better and, more importantly, much bigger CCD detectors.
The larger field size, culminating in the mosaic CCDs with field sizes
up to a square degree that are now
being built at most modern observatories, makes obtaining 
deep multi-color photometry of a large sample of galaxies in an 
efficient manner possible.
Another recent major thrust for the photometric redshift technique
 is the availability of
the  Hubble Deep Field (HDF, Williams et al., 1996) which contains
multi-color data of galaxies to the depth of 29  mag, much too faint
for conventional spectroscopy.

The modern photometric redshift techniques can be divided into two
major approaches.
One is the more traditional method of fitting model galaxy SEDs
to the photometric data;
the other is the newer technique of using a spectroscopic redshift sample as
training set for the photometric redshift sample. 

It should be noted that regardless of which technique is used, the
efficacy of the method depends on the spectral type of the galaxy and the
wavelength region covered by the observed wave bands.
The primary redshift signal from photometric data arises from prominent
breaks in the galaxy SED, such as the 4000\AA/Balmer breaks, and the
Lyman break.
The slope of the photometric measurements on either side of the break
produces additional refinements in the redshift estimate.
The overall continuum curvature of the SED, along with the break size, 
provides information on the spectral type.
Hence, it is expected that galaxies with large spectral
breaks which are straddled by the filter bands (preferably with at least
one band completely on each side of the break) will have the most
robust and accurate measurements. For example, 
for data solely in the observed optical bands, early-type galaxies at
low redshift ($z<0.8$, beyond which the 4000\AA~break enters the $I$ band),
or star-bursting galaxies at high redshift ($z>3$, where the Lyman break
enters the $U$ band)
will have the strongest redshift signal.
On the other hand,
low-redshift star-forming galaxies or any galaxies at $z$ between 1 and 3
will have more uncertain redshift signatures.

\subsubsection{The SED Fitting Method} 
~

\smallskip
All of the early work on photometric redshifts
(Baum 1963; Koo 1985; and Loh \& Spillar 1986) essentially
used the SED template fitting method,
with Loh \& Spillar being the first to explicitly fit SEDs of various
types of galaxies to the data by minimizing $\chi^2$.
The SED fitting technique is necessary if there is no training set of 
redshift data to derive empirically the relationship between
colors and redshift.
Although the method has a certain simplicity in the way it is
applied, a major uncertainty arises from not knowing the precise
template set to be used.
In general, two approaches have been taken.
The first is to use
empirical spectro-photometric SEDs obtained from relatively low 
redshift galaxies, with the most often used templates being  
the SEDs of different spectral types in Coleman, Wu \& Weedman (1990,
hereafter CWW).
An alternative is to use theoretical SEDs from population synthesis
models, such as, for example, the GISSEL models of 
Bruzual \& Charlot (1993, 1996).

A great impetus for using this technique was provided by the HDF
data, as the great depth of these photometric data excludes the
possibility of obtaining a fair spectroscopic redshift
 sample  fainter than $\sim 24$ mag
as the training set.
At high redshift, complications arise due to 
the lack of suitable high quality empirical SED data at short wavelengths,
as the UV spectrum is shifted into the optical band.
Furthermore, high redshift galaxies may have significantly different
SEDs from those of local galaxies even if we are able to obtain
good UV SEDs of local galaxies.
This has led to the use of hybrid templates combining local empirical
SEDs with model SEDs at the short wavelength regime (e.g., see SLY97).
A further complication is the attenuation of the UV spectrum arising from
intergalactic hydrogen.
Although most practitioners statistically correct for this absorption 
(mostly using the method of Madau 1995), the actual amount of
absorption in any line of sight is stochastic, introducing additional
uncertainties.

A large number of groups (for a partial list, see Hogg et al. 1998) 
have derived photometric redshifts for galaxies in the HDF
to as faint as 27 to 28 mag with redshifts running as high as $z=6$. 
The fitting methods used include minimizing $\chi^2$ (e.g., SLY97;   
 Gywn \& Hartwick 1996) and maximizing likelihood
functions (e.g., Lanzetta, Yahil, and Fernandez-Soto 1996).
These results serve as an
excellent illustration of the dependence of the results on the
choice of SED templates, and
 a more detailed discussion is given in Section 3.2.

\subsubsection{The Empirical Training Set Method} 
~

\smallskip
The empirical photometric redshift technique requires a training set
of data in which both the redshifts and photometry are available.
The method tries to obtain an optimal fit between the photometric
and redshift measurements, and uses this fit to predict the redshift
of objects with only photometric data.
Although this method has the advantage of being empirical, and hence
is not dependent on having exact knowledge of the SEDs of galaxies,
obtaining a proper and sufficiently large
training set is very often expensive observationally.
Furthermore, extrapolating the fit to galaxies with magnitudes or
redshifts outside the ranges of the template set must be done with
great care, as it will produce additional unknown uncertainties.

The most detailed application of such a technique has been carried
out by Connolly et al. (1995).
They used a sample of 254 galaxies with redshifts and 4-band  (basically
$U$$B$$R$$I$) photographic photometry with a magnitude limit
of $B_J\le22.5$.
The correlation between photometry and redshift
was analysed in the multi-dimensional space of the flux measurements.
They reported that
a simple linear regression fit of $z$ as a function of the  magnitudes
from the 4 filters produced a rms dispersion, $\sigma_z$, of 0.057.
A quadratic fit, which requires a total of 17 terms in the
4 filters, reduced $\sigma_z$ to 0.047.
They also showed via simulated data that fitting data that are binned
into redshift bands by an iterative procedure can reduce the 
dispersion to less than 0.02.
They found that additional bandpasses do not add more accuracy.
However, this is most likely due to the fact that the method has been applied
to a relatively narrow range of redshift (0.0 to 0.6).
This group has further verified the accuracy of
their method using  a much smaller number
of objects with CCD photometry (Brunner et al. 1997).

Wang et al. (1998) used a modified version of the Connolly et al. method
to derive photometric redshifts from the HDF.
Their method fits 
the spectroscopic redshift to a linear function in the 3 dimensional
color space, instead of the 4 dimensional flux space.

A very different approach, combining SED fitting and the empirical
training methods, is being taken by the CNOC (Canadian Network
for Observational Cosmology) group (Sawicki et al., 1999),
 and we present a preliminary report on this ongoing work.
The CNOC2 field redshift survey (Yee et al. 1998, 1999) 
contains over 5000 redshifts
of galaxies to $R_c\le21.5$ mag with $z<0.75$; the current
results are based on about 1/4 of the sample.
Because of the large sample size and the availability of five-color CCD 
photometry ($U$$B$$V$$R$$I$) for almost all
objects, the CNOC2 survey is an ideal sample for applying 
and testing the empirical photometric redshift technique.
We use a ``hybrid'' technique which first finds the best fitting SED templates
for the training set, and then trains the template set by adjusting the
templates as a function of SED types and {\it redshift} to fit the 
data optimally. 
This trained template set can then be used to obtain redshifts from other
photometric data.
We note parenthetically that there is a major conceptual difference 
between the CNOC method and that of the
Connolly et al. method: the CNOC method does not depend on the 
apparent magnitude of the galaxy for any of the redshift signal.
Using the flux as part of the redshift signal may produce  subtle biases
in an apparent magnitude limited sample, a conjecture which we will
be able to check using the large CNOC2 database.

Using this method we are able to obtain preliminary results with $\sigma_z$
of $\sim0.06$ for objects with a signal-to-noise (S/N) ratio better than 10.
Figure 1 plots $\Delta z$ between spectroscopic and photometric
redshifts as a function of $R_c$ magnitude and redshift for 1387 objects.
The sample can also be divided up into different 
spectral types (via the 5-color photometry).
The results show a worsening of $\sigma_z$ with 
later-type spectra (as expected), but no systematics in $\Delta z$
with either magnitude or redshift.

\begin{figure}
\includegraphics{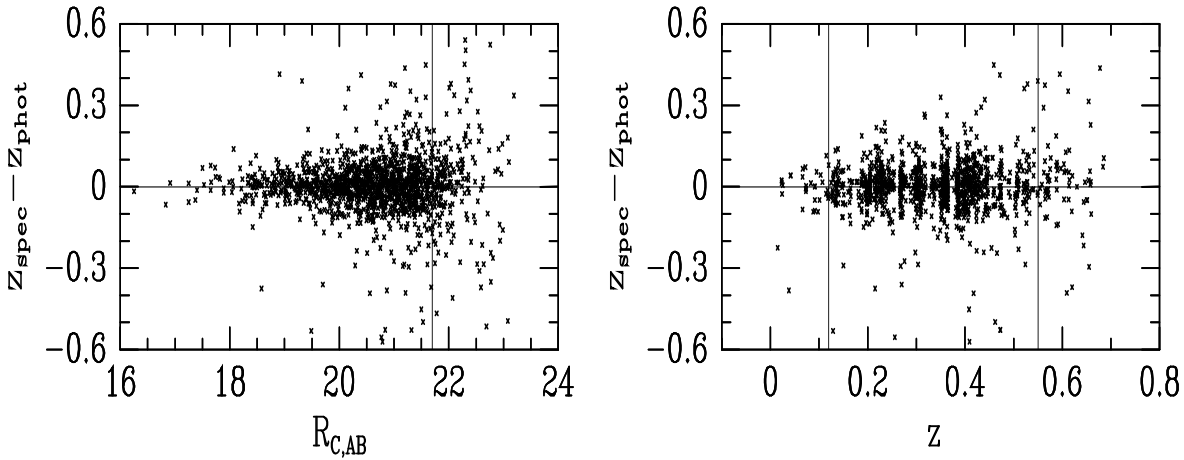}
\vspace{7.0cm}
\caption{$\Delta z$ vs $R$ magnitude and $z$ for 1387 galaxies
in the CNOC2 Field Galaxy Survey sample used as training set.
The vertical lines indicate the nominal spectroscopic limit in $R_{C,AB}$
and the fair sample redshift range.
The rms of $\Delta z$ is about 0.07 for the whole sample.
Note the lack of systematic trend with either magnitude or $z$.
}
\end{figure}

\section{Reliability of Photometric Redshifts}

\subsection {Redshift Uncertainty}

The empirical methods of Connolly et al. (1995) and Sawicki et al. (1999)
show that with 4 or more filters and a S/N ratio of $\sim$10,
an accuracy of $\sigma_z\sim0.05$ can be achieved over the redshift
range of 0 to 0.8.
In the case of the CNOC2 data set (Sawicki et al. 1999), the
sample is large enough that one can actually use a subsample to 
train the templates, and  then test the templates on the 
remaining large set of galaxies with known redshifts. 
Similarly, various authors have shown that comparing their 
photometric redshifts from the HDF obtained by SED fitting with
available spectroscopic redshifts indicates typical agreements of
$\Delta z\sim 0.1$. 
(A detailed discussion of the robustness of the HDF photometric redshifts
will be given in Section 3.2.)

Although direct statistical comparisons of photometric 
and spectroscopic redshifts appear to produce satisfactory  
results, there is  another important
 type of redshift errors that must be  assessed.
This is  termed by Yee, Ellingson, \& Carlberg (1996) as
``catastrophic errors'' in their discussion of errors in
spectroscopic redshift surveys.  This type of errors 
could arise from the misidentification of either emission or
absorption line features, or in the the case of using cross-correlation
techniques, of the cross-correlation peak.
The same problem exists for photometric redshift techniques.
A catastrophic error in photometric redshift determination
occurs when two templates
with very different redshifts occupy regions in the multi-color space
that are too close together, so that for a variety of reasons, the
template with the incorrect redshift is chosen.
This could be the result of poor signal-to-noise ratio in the data,
unrealistic templates being used, or simply a real degeneracy arising from
the very coarse spectral information available within both the data
and the templates.
(For a detailed description of these effects, see SLY97.)
The simplest example of catastrophic error in a sample spanning a
sufficiently large redshift range (as is the case for the HDF data set) 
arises from the confusion of the Lyman break  with the 4000\AA~break.
It is instructive to peruse the redshift likelihood functions plotted in
Lanzetta et al. (1996) in which one very often sees two redshift
peaks, corresponding to the signals from the two strongest
breaks in most galaxy spectra.
Picking the correct redshift has the same connotation as picking
the correct cross-correlation peak in a spectroscopic redshift
analysis.

Catastrophic errors are difficult to assess and catch, and, as we will
show in the next Section, can have severe effects on the scientific
analysis of the data.
A possible way to alleviate this problem is to obtain a larger wavelength
coverage; for example, extending into the near-IR bands.
The larger coverage will either provide significantly more information
on the continuum curvature on both side of a break, or, in some cases,
sample more than one break simultaneously, considerably reducing the ambiguity.
We note that Fernandez-Soto et al. (1998) and
Connolly et al. (1997) supplemented the optical HDF data
with ground-based IR data to improve their redshift determination.

\subsection{The Hubble Deep Field as a Case Study}

The publicly available HDF data instigated a major 
revival of interest in photometric redshifts.
The very deep images in 4 bands provide an excellent data
set for applying the technique to unprecedented depth and
redshift, allowing us to study galaxy evolution over 90\% of the
age of the Universe.
At this point, very few scientific applications of redshift
catalogs from photometric redshift samples have been published
other than those based on the HDF.

Several groups created photometric redshift
catalogs within half a year of the release of the data
(Gwyn \& Hartwick 1996, Mobasher et al. 1996, Lanzetta
et al. 1996, and SLY97), and more followed in the subsequent year
 (e.g., Connolly et al. 1997, Miralles \& Pell\'o 1998; Wang et al. 1998).
This is the only data set which has been analysed independently
for photometric redshifts by a large number of groups, 
offering us the opportunity to compare results and  obtain
some idea on the robustness of photometric redshift determinations.

Here, we can make two kinds of comparisons.
One is the accuracy of the redshift determination as compared
with the increasingly large sample of objects in the HDF with spectroscopically
measured redshifts (most, if not all, from the Keck Telescope).
This comparison would apply only to the set of objects that have been
spectroscopically measured.
Most of the investigations mentioned above have plotted photometric-$z$
versus spectroscopic-$z$ in their paper, and most showed excellent
agreements, with typical spread of $\Delta z\sim0.1$, and 
also a small number of catastrophic redshift errors.
Hogg et al. (1998) specifically conducted a blind test of photometric
redshifts versus a small magnitude-limited spectroscopic redshift
 sample with objects having mostly $z<1.5$.
Five different groups participated in the test, and again, the
results showed that $\Delta z\sim0.1$ were obtained by all groups.

It should be noted, however, that this basic agreement (disregarding the
catastrophic errors, which, as it turned out, could also be due
to erroneous redshift determinations in the spectroscopic sample; see
Lanzetta et al. 1997b, and Sawicki \& Yee 1998) applies only to
the spectroscopic sample, which may not be a homogeneous or fair
magnitude-selected sample.
First, the spectroscopic sample is based on galaxies
that are mostly $<24.5$ mag, whereas the photometric samples extend
the magnitude range considerably, to as faint as 28 mag.
Second, the $z>2$ spectroscopic sample is almost exclusively comprised of
objects preselected to be Lyman-break galaxies based on colors, which
are also galaxies with the largest photometric redshift signals at these
redshifts.
The population of fainter, possibly non-star bursting, $z>2$ galaxies 
are not represented in these direct comparisons.
Furthermore, there is a dearth of spectroscopic redshifts in the
redshift range of $1<z<2$, simply due to the difficulties in measuring
redshifts in this range with an optical spectrograph.
Hence, the apparent agreements between photometric and spectroscopic
redshifts may be over-stated.

An equally important comparison is that between the photometric
redshifts for all the objects obtained by the different groups.
In the case for the HDF data where SED fitting is the
dominant technique used, the primary cause for different results
amongst the various groups is in the details of the template sets used.
Here, we examine the effects of the template sets, and their consequence on
the interpretation.

Of the published photometric redshift works on the HDF from which
we can make direct comparisons of the redshift distributions or derived
luminosity functions, there are basically
two variations of template sets: population synthesis models (e.g., Gwyn \&
Hartwick 1996; Mobasher et al. 1996; and  Miralles \& Pell\'o 1998)
and observed SEDs with UV extensions based on 
synthetic SEDs (e.g., SLY97, and Lanzetta et al. 1996).
The other major difference amongst the various groups is that SLY97 
and Lanzetta et al. included corrections for intergalactic UV absorption, 
while the other 3 did not.
(However, we note that the Gwyn and Mobasher groups have since modified their
methods, see Hogg et al. 1998.)

The different template sets produce significantly different redshift
distributions.  
Gwyn \& Hartwick, Mobasher et al., and Miralles \& Pell\'o
produced similar redshift distributions which have two peaks: one at
$z\sim0.5$ and another more prominent peak at $z\sim2$; while
SLY97 and Lanzetta et al. derived distributions with a redshift peak
near $z\sim0.8$, and a slow decline out to $z\sim4$.
SLY97 was able to reproduce the double-peak distribution by
using simply a pure GISSEL template set.
This is shown in Figure 2.
Hence, it is clear that the precise manner in which the templates
are chosen can have a profound effect on the redshift distribution.
This difference arises from  a large number of catastrophic redshift errors,
most likely in the pure GISSEL model results.
SLY97 plotted the comparison of photometric redshifts obtained using
the extended CWW templates plus Lyman absorption and pure GISSEL
models, showing that catastrophic discrepancies (of up to $\Delta z\sim1$)
primarily occur at
$z>0.8$, where the 4000\AA~break would be in the reddest filter (8140\AA)
 without an anchoring filter further to the red.
A confusion between the 4000\AA~break and the much weaker break
at $\sim2600-2800$\AA~would produce this
aliasing of $\Delta z$ of about 1.
Comparing the redshift catalogs from SLY97, Gwyn \& Hartwick (1996),
and Lanzetta et al. (1996) directly shows discrepancies of similar
magnitudes for objects with  $z>1$ (Sawicki, private communication).
We note that the redshift distribution of Wang et al. (1998), derived
using an empirical fitting of colors, is in basic agreement with those
of SLY97 except for the redshift range between 1 to 2, where there
are the greatest number of catastrophic discrepancies.
Hence, one can conclude that in specific redshift intervals, where
there is  confusing, or little or no redshift information in the
observed wave bands, redshift determination may contain large 
(and likely systematic) uncertainties and the results  may depend
sensitively on the templates used.

\begin{figure}
\includegraphics{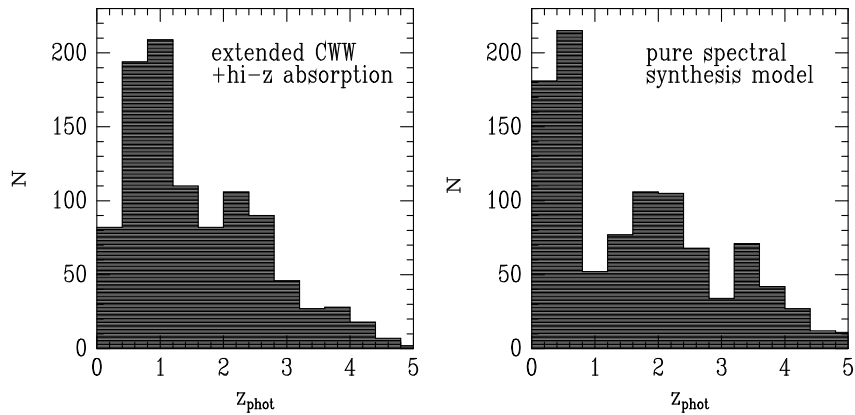}
\vspace{6.5cm}
\caption{Comparison of photometric  redshift distributions of galaxies
brighter than $F814W=27$ mag from the HDF.
The left  panel shows the distribution obtained by using CWW SEDs
extended with GISSEL models with intergalactic absorption, the right
panel shows the distribution obtained by using pure GISSEL models which
are less realistic.
}
\end{figure}

The different redshift distributions can produce very different 
interpretations of the star formation history of the Universe.
This is most clearly seen in the luminosity function  (LF) of galaxies
at high redshift derived by the three groups who have performed
the analysis.
Gwyn \& Hartwick (1996) and Mobasher et al. (1996) produced similar
LFs at $z>2$ with are considerably flatter at both the bright
and faint ends with significant excess density over those of SLY97.
As a consequence, the amount of luminosity evolution estimated
 by the two groups differs from that of SLY97 by a factor of several.

\section{Other Applications}
\subsection {The Roles of Photometric Redshifts in a Spectroscopic Redshift
Survey}

The redshift information contained in multi-color data can be extremely
useful in a spectroscopic redshift survey.
For surveys which target objects of specific interest or redshift
range, (e.g., members of a galaxy cluster), this information can be
used to preselect galaxies for spectroscopic observation.
The most spectacularly successful example of this is the systematic
survey of $z>3$ Lyman-break galaxies by Steidel and his collaborators
(e.g., Steidel et al. 1996) using $U$-band drop-outs.
In their survey, images in 3 bands are used to select objects which
are detected in the two redder bands, but extremely faint in the bluest
band.
The drop out in the $U$ band is due to the Lyman-break at a redshift of
greater than $\sim 3$.
The sky density of these galaxies is typically 10$^3$ per square
degree at $R\sim25$, or about 1\% of all the objects.
Hence, a preselection of targets based on colors makes such a survey
possible.

Another useful function of photometric redshift information is
in verifying and improving the redshift determination in a
general redshift survey.
As mentioned above, spectroscopic redshifts suffer from catastrophic
errors when a misidentification of features or the correlation peak occurs.
Photometric redshift information, either in the form of photometric
redshift or simple colors, can be used to verify the redshift.
In a large redshift survey, color information can be used for this
purpose in the data reduction pipeline, flagging possible redshift errors
that may require human intervention.
We note that several HDF redshifts have been found to be in error
based on discrepancies with the photometric redshifts (see
Lanzetta et al. 1997b and Sawicki \& Yee 1998).
A second application of photometric redshifts in a general redshift
survey is using them to
assist in recovering spectroscopic redshifts from marginal S/N 
ratio spectra.
In a low S/N ratio spectrum, the true correlation peak often has to compete
with other noise spikes due to beating or confusion between real features
and features produced by noise or bad sky subtraction.
The photometric redshift can be used to narrow down the possible redshift range
of the object, allowing the search for a significant correlation peak
over a smaller redshift space.
In the CNOC2 redshift survey with its extensive 5-color database, 
the photometric redshift information is used both to verify
measured redshifts
and to extend the redshift catalog (Yee et al. 1999)

\subsection {The Red Sequence in Galaxy Clusters as a Redshift Indicator}

In general, photometric data containing only 2 bands (i.e., one color)
are of limited use in extracting redshift information.
This is due to the degeneracy in the dependence of a single
color on redshift and spectral type.
However, in special cases where the morphological type of galaxies
are specified, e.g., in the core of galaxy clusters where almost all
galaxies are expected to have an early-type spectrum, a single
color can provide an extremely accurate estimate of the redshift.

The early-type galaxies in clusters form a color-magnitude relation (CMR)
on the color-magnitude diagram with brighter galaxies being redder.
These galaxies are also the reddest objects in the cluster -- hence
the name ``red sequence''.
From the photometric data of
 a sample of 45 Abell clusters with redshift between 0.04 and 0.18,
L\'opez-Cruz (1997) shows that the $B-R$ colors of the red sequence
at a fixed apparent $R$ magnitude can be fitted by a quadratic function
to $z$ with a rms disperson of less 0.008 in $z$, considerably better
than any photometric redshift methods tested so far.
The above example is equivalent to the method of using 
a known redshift sample as a training set.
In the case where such a training set does not exist, spectral synthesis
models of the CMR, such as those by Kodama \& Arimoto (1997),  can be used.

Gladders \& Yee (1999) have explored using the red sequence method as
an inexpensive way of finding clusters/groups from two-band imaging
data.
The algorithm produces the surface density of galaxies on the sky using
galaxies in slices of model CMR in the color-magnitude plane.
The color slices correspond to redshift slices, based on models from
Kodama \& Arimoto (1997).
Figure 3 illustrates the result using CNOC2 data as
gray scales of galaxy surface density
at selected redshift.
A number of density peaks are apparent.
There are two points of note.
First, the significance of the density peaks in general increases when
viewed in its redshift slice. 
Second, one of the peaks in the overall density map breaks up into two
peaks at the same position but different redshifts;
hence, the redshift slicing by the CMR colors is a very effective
in alleviating the problem of line-of-sight projection in searches
for groups and clusters.
These results have been verified by the spectroscopic redshift data.
We also note that these peaks correspond to concentrations that are
considerably poorer than Abell Richness Class 0 clusters, making this
a very promising one method for finding galaxy groups and clusters. 

\begin{figure}
\includegraphics{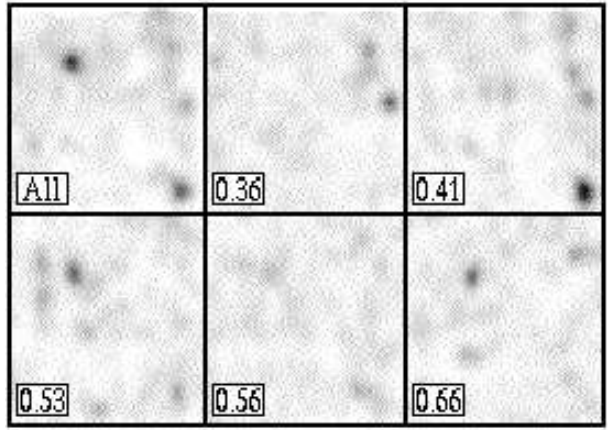}
\vspace{6.0cm}
\caption{
Gray scale plots of galaxy surface density in a $\sim27\times24$ 
arcmin field from the CNOC2 survey.
The ``All'' panel shows the density of all the data in $R$ band.
The other panels show the density in selected CMR color slices
with the corresponding redshift shown on the lower left corner.
Note the increase in the significance of the peaks in the
redshift slice map, and the breaking up of the most prominent peak
into  two peaks in redshift space.
}
\end{figure}

\section{Summary}

Photometric redshifts have undergone a great revival of interest in the
past few years.
This is in part due to the developments in new detectors and new techniques,
and in part due to the availability of the very deep HDF data.
Both major approaches to photometric redshifts -- the use of  SED fitting or 
empirical training sets -- have seen considerable advances, and the methods
have been applied to very large galaxy samples (e.g., the CNOC2 survey), and
galaxies out to large redshifts (e.g., the HDF).
Current evidence suggests that typical {\it  statistical}
accuracy of $\Delta z$ of 0.05 to 0.10 can be expected.
However, this conclusion is strictly valid only over 
redshift intervals in which the photometric data contain
significant and unambiguous redshift signals.
There are two situations which may result in 
``catastrophic'' redshift errors, and hence must be dealt
with great care.
These are: confusion of spectral breaks which provide most of the
redshift signals and the lack of strong features in the SED within
the wavelength range of the data.
It is interesting to note these are the very same problems encountered
in spectroscopic redshift determinations.
Photometric redshift information is also contained in data with two or
three filters. 
These can be used to estimate redshifts of objects with specific properties,
such as Lyman-break galaxies and red galaxies in clusters.
Photometric redshifts can also be used to verify and improve spectroscopic
redshift determination.

In summary, the technique of photometric redshift is an 
excellent tool with many important applications.
However, the accuracy and robustness of the method depend on many factors,
and great care must be taken to ensure that systematic effects and
catastrophic redshift errors are minimized and understood.

\bigskip
I would like to thank Marcin Sawicki and Mike Gladders for providing
the graphics for the paper and insightful discussions, and also
Erica Ellingson for useful comments.


%
\vfill
\end{document}